\documentclass[12pt,epsf]{article}
\newcommand{\eps}{\epsilon}

\newcommand{\EE}{{\cal E}}

\newcommand{\LL}{{\cal L}}

\newcommand{\TT}{{\cal T}}

\newcommand{\PP}{{\cal P}}

\newcommand{\be}{\begin{equation}}
\newcommand{\ee}{\end{equation}}
\newcommand{\ben}{\begin{eqnarray}\displaystyle}
\newcommand{\een}{\end{eqnarray}}
\newcommand{\refb}[1]{(\ref{#1})}
\newcommand{\p}{\partial}

\newcommand{\al}{\alpha}

\begin{document}

{}~ \hfill\vbox{\hbox{hep-th/0205098}
\hbox{UUITP-04/02}}\break

\vskip 1.cm

\centerline{\large \bf Rolling the tachyon in super BSFT}
\vspace*{1.5ex}

\vspace*{4.0ex}

\centerline{\large \rm Joseph A. Minahan\footnote{E-mail:
joseph.minahan@teorfys.uu.se}}
\vspace*{2.5ex}
\centerline{\large \it Department of Theoretical Physics}
\centerline{\large \it Box 803, SE-751 08 Uppsala, Sweden}
\vspace*{3.0ex}

\vspace*{4.5ex}
\medskip
\bigskip\bigskip
\centerline {\bf Abstract}
We investigate the rolling of the tachyon on the unstable D9 brane
in Type IIA string theory by studying the BSFT action.  The action is
known for linear profiles of the tachyon, which is the expected asymptotic
behavior of the tachyon as it approaches the closed string vacuum, as 
recently described by Sen.  We 
find that the action does indeed seem consistent with the general
Sen description, in that it implies a constant energy density with
diminishing pressure.
However, the details are somewhat different from an effective field theory 
of Born-Infeld type.  For instance, the BSFT action implies there are
poles for certain rolling velocities, while a Born-Infeld action would have
a cut.  We also find that solutions with pressure diminishing from either
the positive or negative side are possible.

\bigskip

\vfill \eject
\baselineskip=17pt


There has been much recent interest in time dependent solutions of string
theory.  Among other things, these solutions could have interesting cosmological implications.   Not surprisingly, time dependent solutions in the closed
string sector are much more difficult to handle, because of the presence
of gravity.  However, one expects that time dependent solutions involving
the open string sector can be more easily controlled, since at tree level,
gravity is turned off.

The first such open string solutions were the spatial branes in
\cite{0202210}.  Here, the tachyon has a time dependent profile with the tachyon
rolling up and down the potential.  It was assumed that the tachyon is
damped as it rolls down the hill, with the energy radiating into closed
string modes, so that the tachyon stops at the bottom of the potential.

This then raises a puzzle about what happens at tree-level, where radiation
into closed string modes does not occur.  Energy should be conserved,
but there are no open string modes where the energy can be radiated into.  
On the other,
hand, if the energy is not being lost, it would seem that
the tachyon configuration would be oscillating back and forth through the
closed string vacuum.  But there is a strong feeling that
there is only one direction in field space 
out of the closed string vacuum, so it seems impossible to set up such 
a ringing
configuration.

These puzzles were recently answered by Sen \cite{0203211,0203265,0204143}, 
who argued that while the
tachyon rolls down the potential, conserving energy, it never reaches the closed string vacuum.  Hence, there is no need for the tachyon to roll up on the
other side of the closed string vacuum.

Sen described an effective field theory that exhibits this behavior.  Consider
the Born-Infeld type action \cite{0003122,0003221,0004106}
\be\label{action}
S=-\int d^Dx\sqrt{1-\al'\p_\mu T \p^\nu T}V(T),
\ee
where $V(T)$ is the tachyon potential that approaches $0$ as $T\to\infty$.
If we consider spatially homogenous solutions to the equation of motion,
we find the conserved energy density is given by
\be
\EE=\frac{V(T)}{\sqrt{1-\dot T^2}}.
\ee
Hence, energy conservation requires that $\dot T\to1$ as $T\to\infty$, hence
the field configuration approaches a constant ``velocity''.   On the other
hand, the pressure is given by
\be\label{pressure}
\PP=-V(T)\sqrt{1-\dot T^2},
\ee
and so  goes to zero as $T\to\infty$.  Thus, as the brane decays, we
are left with a pressure tachyon matter whose energy density depends on
initial conditions.  Cosmological considerations for this and other tachyon
actions have
been considered in \cite{0204008}--\cite{0107058}.

Suppose that we consider specific tachyon profiles in the action in
\refb{action}, which do not necessarily correspond to solutions to
the equations of
motion.  For instance, suppose that we consider a tachyon profile with
\be
T=ut.
\ee
Assuming that the potential is well enough behaved, the action 
becomes
\be\label{BIu}
S(u)=-\ \frac{\sqrt{1-u^2}}{u}V_0\int dT V(T)=-\ \frac{\sqrt{1-u^2}}{u}V_0 K
\ee
where $V_0$ is a volume factor and $K$ is independent of $u$.   
Hence, if the effective field theory
is of Born-Infeld type, we expect to find a branch cut in $S(u)$ at 
$u^2=1$.

In boundary string field theory (BSFT), the effective actions for
tachyon fields were computed for simple profiles \cite{0009103,0009148}.  In the bosonic theory,
the profiles had the form \cite{9208027,0009103,0009148}
\be
T=a+u_ix_i^2,
\ee
where the $x_i$ are spatial coordinates.  In particular, it was shown that
the condensation of a D25 brane to a D(25-$p$) corresponds to $u_i=0$,
$i=1,..,25-p$ and $u_i\to\infty$ for $i=26-$p$,..25$.  Unfortunately
for our purposes here,  the tachyon field has to be quadratic, or independent
of the $X_i$ since the tachyon potential is unbounded below as $T\to-\infty$.
Hence, the integral in \refb{BIu} would not be convergent.

Instead, let us consider the BSFT for the type IIA D9 brane.  This
was considered in \cite{0010108}, 
where the field theory action was conjectured to 
be
\be\label{BSFTpost}
S=Z,
\ee
where $Z$ is the partition function for a boundary perturbation of
a free conformal field theory.  This conjecture was later verified
in \cite{0103089,0103102} using a BV formulation.    Let the tachyon profile have 
the form
\be
T=u\ x
\ee
where $x$ is one spatial direction.  Let us define $y=\al'u^2/2$.
In \cite{0010108}, it was
shown that assuming  \refb{BSFTpost},
the Euclidean action for such a profile is given by
\be\label{Sy}
S(y)=Z(y)=\TT_9V_0\sqrt{\al'}\ 4^y\ 
\frac{\sqrt{y}\ \Gamma(y)^2}{2\ \Gamma(2y)},
\ee
where $\TT_9$ is the tension of the unstable brane and $V_0$ is the 9 
dimensional volume factor.
This action should be thought of as an action with the fields restricted as
follows:
\begin{eqnarray}\label{action2}
S(y)&=&\TT_9\int d^{10}x\ \LL(T,T',T'',...)\Bigg|_{T'=u,T''=0..}\nonumber\\
&=&\TT_9\sqrt{\frac{\al'}{2y}}\int d^9x\ dT\LL(T,\sqrt{\frac{2y}{\al'}},0,...)
\end{eqnarray}
The authors in \cite{0010108} then argued that the stable D8 brane
is the profile with $y\to\infty$ and they obtained the correct ratio of 
tensions between the unstable D9 brane and the stable 
D8 brane.

Of course, what we really want is the Lagrangian and not just the action.
It is known that for $\dot T=0$, the Lagrangian is \cite{0010108,0009246}
\be
-\TT_9\ e^{-\frac{1}{4}T^2}.
\ee
We will now make the further assumption that the Lagrangian, up to first 
derivatives, is given by
\be
-\TT_9 \int dt d^9x\ f(\al'\p_\mu T\p^\mu T/2)e^{-\frac{1}{4}T^2}.
\ee
Comparing this expression with \refb{Sy}, we see that
\be
f(y)=4^{y-1/2}\ \frac{y\ \Gamma(y)^2}{\sqrt{2\pi}\ \Gamma(2y)}.
\ee
Note that for large $y$,
\be
f(y)\approx \frac{1}{2}\sqrt{\al'\p_\mu T\p^\mu T},
\ee
which does have Born-Infeld like behavior \cite{0011226}.

Let us now Wick rotate the spatial direction into a time direction.  Under
the Wick rotation, $x\to i\ t$ and $u\to -i\ u$, so that the profile is
\be
T=u\ t.
\ee
But we also see that $y\to -y$ and  the action rotates to $S\to -iS$,
with the Minkowskian action given by
\be
S=\TT_9V_0\sqrt{\al'}\ 4^{-y}\ \frac{\sqrt{y}\ \Gamma(-y)^2}{\sqrt{2}\ \Gamma(-2y)}.
\ee
Hence, the action has poles for $y$ any nonnegative integer.

The pole at $y=0$ is already in the Euclidean action and corresponds to
the infinite volume factor between the D9 brane and the D8 brane.
Let us then consider the other poles.  If we write $y=n-\eps$, then
the action is approximately
\be
S(y)\approx \TT_9V_0\sqrt{\al'}\ 4^{-n-1/2}\ 
\frac{\sqrt{n}\ \Gamma(2n+1)}{\Gamma(n+1)^2}\ 
\frac{1}{\eps}.
\ee
Hence, we may infer that for profiles near these poles, the action
is approximately
\be\label{poleaction}
S\approx\TT_9\int dt d^9x \frac{V_n(T)}{2n-\al'\p_\mu T\p^\mu T},
\ee
where $V_n(T)$ is
\be
V_n(T)=4^{-n-1/2}\ \frac{\sqrt{\al'}\ \Gamma(2n+1)}{\sqrt{2\pi}\ \Gamma(n)\Gamma(n+1)}\ e^{-T^2/4}.
\ee

For $\al'\dot T^2$ close to $2n$, the action will be dominated by the term
in \refb{poleaction}.  For such a term, the energy density for a rolling
tachyon is given by
\be\label{newenergy}
\EE\approx\TT_9\ \frac{3\al'\ \dot T^2-2n}{(2n-\al'\ \dot T^2)^2}V_n(T),
\ee
and the pressure is
\be\label{newpressure}
\PP\approx\TT_9\ \frac{1}{(2n-\al'\ \dot T^2)}V_n(T).
\ee

Let us now consider the consequences of \refb{newenergy} and 
\refb{newpressure}.  First we see that the BSFT action is consistent with the 
general Sen picture
that the rolling tachyon asympotically reaches a constant velocity with 
conserved energy.  Since $V_n(T)$  falls off rapidly to zero as $T\to\infty$,
$\al'\ \dot T^2$ 
must rapidly approach $2n$, for some positive integer $n$. 
Second, in order for the energy density to remain constant,
the pressure must go to zero as $T\to\infty$.

However, there are some interesting differences as compared to the Born-Infeld
action.  First, there is no longer a maximum rolling velocity, it is now
possible to have $\al'\ \dot T^2>2$.  However, it takes an infinite amount of
energy to cross from $\al'\ \dot T^2<2n$ to $\al'\ \dot T^2>2n$.   
So if the initial
conditions have $2n<\al'\ \dot T^2<2(n+1)$, the tachyon rolling velocity
will stay within that window.

The second difference between this behavior and the Born-Infeld model is
that while the pressure approaches 0 in both cases, in the Born-Infeld case,
it is always negative.  However, the pressure in \refb{newpressure}
can be either negative or positive.
So for example, if the initial conditions start with $\dot T=0$, the pressure
starts out negative, equalling the negative of the energy density, but then
crosses through zero pressure, before relaxing back down to zero.

However, it is also possible to approach zero pressure from the negative
direction.  For instance, if $\al'\ \dot T^2$ is slightly greater than
$2n$, then the rolling velocity will relax down to $\al'\ \dot T^2=2n$.
Since $\al'\ \dot T^2>2n$, the pressure is negative.  However, if 
$\al'\ \dot T^2$ is too big, the velocity will increase toward 
$\al'\ \dot T^2=2(n+1)$,
and the pressure will be positive.

Note that positive pressures should be possible.  Indeed, in Sen's
analysis the static vacuum can be perturbed by the marginal operator
\be
\lambda \sinh(2t/\sqrt{\al'}),
\ee
in which case, the pressure is proportional to 
\be
\PP\sim 1-\frac{1}{1+e^{2t/\sqrt{\al'}}\sinh^2(\lambda\pi)}
-\frac{1}{1+e^{-2t/\sqrt{\al'}}\sinh^2(\lambda\pi)}.
\ee
This perturbation corresponds to having an initial rolling velocity at
the height of the potential at $t=0$.  If $\sinh(\lambda\pi)>1$, then
the pressure is positive. 

Of course, there are many assumptions that go into this analysis.  
The first is that the string field action does have the form in
\refb{BSFTpost}.  The second assumption is that the poles that appear
in the action in \refb{action2} are actually due to poles in the integrand,
that is the Lagrangian, as opposed to poles that arise after integrating over $T$.  
For example,
the action would have a pole if the Lagrangian had a term of the form
\be
\exp(-T^2(2n-\al'\ \dot T^2)^2/16n^2).
\ee
For such a term, the constant energy solutions would still have
$\dot T$ relaxing to a constant.  However, the asymptotic pressure
would not be zero.

Finally, terms with higher derivatives of $T$ do not contribute to
the action if $T$ has a 
linear profile.  The appearance of such terms in the full lagrangian
could effect the physics.

\medskip
\noindent{\bf Note added:}  After this paper was completed
 we received \cite{0205085}, which has significant overlap with
this work.

\bigskip
\noindent {\bf Acknowledgments}:
I would like to thank the CTP at MIT
 for hospitality during the course of 
this work.   I also thank Barton Zwiebach for many helpful conversations
and for early collaboration on this work.  This research was
supported in part by the NFR.

\end{document}